\documentclass[12pt,times,authoryear]{article}

\usepackage{amssymb,amsfonts,amssymb,subcaption,amsmath,caption,mathtools,float,hyperref,graphicx}
\usepackage{authblk}
\usepackage[round]{natbib}
\newcommand{\Bernoulli}{{\text{Bernoulli}}}

\newcommand{\Betadist}{{\text{Beta}}}
\newcommand{\betadist}{{\text{Beta}}}

\newcommand{\CRT}{{\text{CRT}}}

\newcommand{\Dirichlet}{{\text{Dirichlet}}}

\newcommand{\Gammadens}{{\text{Gamma}}}
\newcommand{\gammadens}{{\text{Gamma}}}
\newcommand{\Gammadist}{{\text{Gamma}}}
\newcommand{\gammadist}{{\text{Gamma}}}
\newcommand{\Poisson}{{\text{Poisson}}}

\newcommand{\Multinomial}{{\text{Multinomial}}}

\newcommand{\normaldist}{{\text{Normal}}}
\newcommand{\PG}{{\text{PG}}}

\makeatletter
\newcommand*\bigcdot{\mathpalette\bigcdot@{.5}}
\newcommand*\bigcdot@[2]{\mathbin{\vcenter{\hbox{\scalebox{#2}{$\m@th#1\bullet$}}}}}
\makeatother

\title{Hierarchical Topic Presence Models}
\author[1]{Jason Wang}
\author[1]{Robert E. Weiss}
\affil[1]{Department of Biostatistics,
	Fielding School of Public Health\\
	University of California Los Angeles}

\begin{document}
\maketitle

\begin{abstract}
Topic models analyze text from a set of documents. Documents are modeled as a mixture of topics, with topics defined as probability distributions on words. Inferences of interest include the most probable topics and characterization of a topic by inspecting the topic's highest probability words. Motivated by a data set of web pages (documents) nested in web sites, we extend the Poisson factor analysis topic model to hierarchical topic presence models for analyzing text from documents nested in known groups. We incorporate an unknown binary topic presence parameter for each topic at the web site and/or the web page level to allow web sites and/or web pages to be sparse mixtures of topics and we propose logistic regression modeling of topic presence conditional on web site covariates. We introduce local topics into the Poisson factor analysis framework, where each web site has a local topic not found in other web sites. Two data augmentation methods, the Chinese table distribution and P\'{o}lya-Gamma augmentation, aid in constructing our sampler. We analyze text from web pages nested in United States local public health department web sites to abstract topical information and understand national patterns in topic presence.
\end{abstract}

\section{Introduction}
\label{introduction}
Probabilistic topic models have been used to abstract topical information from collections of text documents by modeling documents as a mixture of $K$ latent topics where each topic is itself a mixture of the $V$ unique words in a vocabulary. A topic is characterized by a vector of word probabilities and a document is characterized by a vector of topic probabilities. Topic-word distributions and document-topic distributions describe the prevalence of words in a topic and topics in a document, respectively. Topic models such as latent Dirichlet allocation (LDA) are constructed under a Dirichlet-multinomial framework, where words in a document follow a multinomial distribution with a Dirichlet prior \citep{Blei, rtm}. More recently, \cite{pfa} introduced the Poisson factor analysis (PFA) framework which models word counts with a Poisson likelihood. \cite{nbpmm} demonstrate computational advantages of PFA models over LDA models. We discuss and propose novel topic models in the Poisson factor analysis framework.
\\ \indent
Our work is motivated by a text data set of web pages nested within local health department web sites in the United States. We treat web pages as separate documents nested in web sites. We are interested in identifying health topics covered by health department web sites, how frequently topics are covered and which topics are or are not covered in individual web sites. As is usual in PFA and LDA models, we label topics by inspecting the most frequent words in the topic.
\\ \indent
Some development of models for nested or clustered documents has occurred with LDA. 
Some models address nesting by modeling multiple levels of document-topic distributions \citep{multdoctop}.
Some models explicitly model topics that are unique to documents in a given cluster \citep{localglobal,haltlda}. 
\citet{haltlda} further proposed hierarchical priors on document-topic distributions to accommodate the belief that which topics are more common vary from web site to web site.
In contrast, there has been little development of PFA for nested documents. 
\\ \indent
Different public health department web sites will likely contain different subsets of topics. Some health topics will be present in most web sites while other health topics may be rarer or may be of specific interest depending on demographic or geographic characteristics of the local health department. Thus, we propose modeling topic presence conditional on covariates.
\\ \indent
Sparsity inducing priors model documents as a mixture of a subset of the possible topics and can be implemented by introducing unknown topic presence binary indicators for whether a given topic contributes words to a particular document. Topic presence has previously been introduced in non-nested document collections \citep{ftm, pfa, ftmieee, dpfa}. \citet{pfa} proposed the sparse Gamma-Gamma Poisson factor analysis, also known as the negative binomial focused topic model (NB-FTM) \citep{nbpmm} in the PFA framework. The NB-FTM needs to be adapted for nested documents. Further, prior researchers have not modeled topic presence as functions of document covariates.
\\ \indent
Nesting of web pages in web sites allows for topic presence modeling at the web site level, the web page level, or both. We propose three topic presence models (TPM).
The first TPM has topic presence at the web site level while all web pages are mixtures of the topics present in their respective web sites. The second TPM is for web pages only while web sites are mixtures of all topics. The third model has TPMs for both web site and web page. Web site (web page) topic presence is a vector of unknown binary variables that identifies the subset of topics in a web site (web page) -- topics must be present at the web site level to be present in a web page nested in the web site.
\\ \indent
Previous topic presence models have modeled topic presence as a priori independent where the unknown probabilities of topic presence have fully known priors. We extend this to allow topic presence probabilities to be a priori exchangeable, where we estimate the global mean and variance of the probability of a given topic's presence across web sites (or web pages). We extend this model and consider a logistic regression model for topic presence where probability of topic presence is modeled conditional on covariates.
\\ \indent
When we model multiple web sites, pages of a web site are likely to include common local words or phrases such as names and locations that are not commonly found in the web pages of other web sites. These local words form a \textit{local topic} that is unlikely to be found on other web sites. Local topics are unique to a web site while global topics can be present in multiple web sites.
Local topics have been introduced to LDA models \citep{localglobal,haltlda} and \citet{haltlda} showed that including local topics reduces the number of global topics needed without sacrificing performance.
\\ \indent
In topic models, typically the number of topics $K$ is a parameter to be specified. The number of topics $K$ can be modeled however \cite{nbpmm} suggest that sufficiently large $K$ provides a good approximation to models with $K$ unknown. We take $K$ as a parameter that we tune.
\\ \indent
We derive a Gibbs sampler for inference after suitable data augmentation. We need two families of auxiliary random variables distributed as the Chinese restaurant table (CRT) distribution to sample topic parameters from conditional posterior Gamma distributions \citep{hdp,pfa}. We introduce families of P\'{o}lya-Gamma distributed \citep{polyagamma} auxiliary random variables to allow us to sample our logistic regression coefficients in the topic presence models as Gibbs steps.
\\ \indent
We consider several hierarchical PFA models, with and without local topics. We also consider six topic presence models: at the web site level we consider using covariates to predict presence, an exchangeable prior, and topics always present. At the web page level we consider exchangeable and topics always present. We compare models with perplexity, a measure of predictive fit \citep{Wallach2009,pfa} that we extend to our hierarchical settings. We provide a quick automated approach using the most probable words of a topic to check if our models correctly capture patterns in web site topic presence.
\\ \indent
The next section \ref{poissonfactor} presents the LDA and PFA models in our context then section \ref{nestedtopic} extends PFA to be hierarchical PFA with local topics and hierarchical topic presence models. Section \ref{analyzingwebcontent} presents our analysis of local health department web sites and the paper closes with discussion.

\section{Poisson Factor Analysis}
\label{poissonfactor}
We first present notation for the Poisson factor analysis (PFA) model in the context of our hierarchical data set and extend PFA to include local topics. Let $i = 1, \ldots, M$ index web sites and let $j = 1,\ldots, N_i$ index web pages nested in web sites with $N_i$ web pages in web site $i$.
\\ \indent
For PFA, we treat individual web pages as separate documents. Let $k = 1, \ldots, K$ index global topics where $K$ is set in advance. The vocabulary or set of unique words in a document collection is known and has length $V$ and we let $v = 1, \ldots, V$ index words in the vocabulary. Poisson factor analysis (PFA) models word counts with a Poisson likelihood. Let $z_{ijkv}$ be the latent count of word $v$ from topic $k$ in web page $j$ of web site $i$. Let $\phi_{kv}$ be the probability of word $v$ in topic $k$ and let $\theta_{ijk}$ be the weight of topic $k$ in web page $j$ of web site $i$ such that $\theta_{ijk}$ is the expected count of words from topic $k$ in web page $j$ of web site $i$. Then $\phi_{kv}\theta_{ijk}$ is the expected count of word $v$ from topic $v$ in web page $j$ of web site $i$, and PFA models latent counts $z_{ijkv} | \phi_{kv}, \theta_{ijk} \sim \Poisson(\phi_{kv}\theta_{ijk})$.
\\ \indent
We model one local topic for each web site. Let global topic word probability vectors be $\phi_k = (\phi_{k1}, \ldots, \phi_{kV})'$ and let the local topic word probability vector be $\psi_{i} = (\psi_{i1}, \ldots, \psi_{iV})'$ for web sites $i = 1, \ldots, M$, such that $\psi_{iv} > 0$ and $\sum_{v=1}^{V}\psi_{iv} = 1$. Only web pages of web site $i$ can have non-zero topic weight for local topic $\psi_{i}$. Define $\Phi_{i} = (\phi_1, \ldots, \phi_K, \psi_i )'$ to be the $(K + 1) \times V$ matrix of word probability vectors for all global topics plus web site $i$'s one local topic word probability vector. Then $\Phi_{ikv}$ is the probability of word $v$ in topic $k$ in web site $i$, where $k = 1, \ldots, K + 1$ where $k \le K$ indexes the $K$ global topics while $k = K + 1$ is the local topic for web site $i$. Extend the definitions of $\theta_{ijk}$ and $z_{ijkv}$ to have $k$ run from 1 to $K+1$. The PFA local topic model (PFA-LT) models $z_{ijkv}$ as
\begin{align}
z_{ijkv}|\Phi_{ikv},\theta_{ijk} &\sim \Poisson(\Phi_{ikv}\theta_{ijk}). \label{pfatwo}
\end{align}
From now on, for models with local topics, $k$ runs from 1 to $K + 1$ while for models without local topics, $k$ runs from 1 to $K$.

\section{Poisson Factor Analysis with Local Topics and Hierarchical Topic Presence}
\label{nestedtopic}
Topic presence is a web site or web page binary variable that indicates whether a topic is present or not in the web page or web site. We can model web site topic presence, web page topic presence, both, or neither. Let $b_{ik} = 1$ indicate that topic $k$ is present in web site $i$ and let $c_{ijk} = 1$ indicate that topic $k$ is present in web page $j$ of web site $i$. When $b_{ik}$ and $c_{ijk}$ are both included in our model, topic $k$ is present in web page $j$ of web site $i$ only if both $b_{ik} = c_{ijk} = 1$. The number of words in web page $j$ of web site $i$ is $z_{ij\bigcdot\bigcdot}= \sum_{kv} z_{ijkv}$. We model topic weights $\theta_{ijk} \ge 0$ conditional on global topic weight parameters $r_k$, web site topic presence $b_{ik}$ and web page topic presence $c_{ijk}$ such that
\begin{align}
\theta_{ijk}|r_k, b_{ik}, c_{ijk} &\sim \gammadens(r_{k}b_{ik}c_{ijk}z_{ij\bigcdot\bigcdot}, 1). \label{thetadef}
\end{align}
Thus, $\theta_{ijk} = 0$ with probability 1 if $b_{ik} = 0$ or $c_{ijk} = 0$. The gamma density in \eqref{thetadef} has mean equal to the variance as for smaller $\theta_{ijk}$, we want smaller variance and for larger $\theta_{ijk}$ we want larger variation. The scale parameter in \eqref{thetadef} is 1 as there is an arbitrary scaling involved which is unnecessary for modeling the counts. 
\\ \indent
The number of words $z_{ij\bigcdot\bigcdot}$ is a scaling factor to increase or decrease $\theta_{ijk}$ as for a given $r_k$, web pages with more words will have larger $\theta_{ijk}$ compared to web pages with fewer words. Omitting $z_{ij\bigcdot\bigcdot}$ in \eqref{thetadef} would require a factor indexed by $ij$ to model the web page word count $z_{ij\bigcdot\bigcdot}$. As size $z_{ij\bigcdot\bigcdot}$ is at best an ancillary statistic, we condition on $z_{ij\bigcdot\bigcdot}$ in \eqref{thetadef}. Conditional on the total $z_{ij\bigcdot\bigcdot}$ of a set of $KV$ independent Poisson random variables (PRVs), the set of PRVs are distributed as multinomial. 
However, $KV$ is very large, the probabilities are small and the Poisson approximation to the multinomial distribution will be quite accurate. 
In modeling counts $z_{ijkv}$ as Poisson in \eqref{pfatwo}, we do not directly condition on $z_{ij\bigcdot\bigcdot}$ but only indirectly in \eqref{thetadef}, so the Poisson approximation should be quite acceptable.
\\ \indent
We place a gamma hyperprior on the $r_k$ for $k = 1,\ldots,K+1$,
\begin{align*}
r_k|r_0 &\sim \Gammadens(r_0, 1) \\
r_0 &\sim \Gammadens(d_{r0}, e_{r0})
\end{align*}
and $r_0$ is a prior mean global topic weight with fixed prior hyperparameters $d_{r0}$ and $e_{r0}$. 
We place a Dirichlet prior on word probability vectors $\phi_k$ and $\psi_i$ such that
\begin{align*}
\phi_{k} &\sim \Dirichlet(\alpha_{\phi} 1_V), \\
\psi_{i} &\sim \Dirichlet(\alpha_{\psi} 1_V),
\end{align*}
where $\alpha_{\phi}$ and $\alpha_{\psi}$ are fixed hyperparameters and $1_V$ is a ones vector of length $V$ and the $\Dirichlet(c 1_V)$. 

\subsection{Models for Topic Presence Probabilities}
Web site topic presence $b_{ik}$ is given a Bernoulli($\pi_{ik}$) prior, where $\pi_{ik}$ is the probability of topic $k$ being present in web site $i$. We consider three prior specifications for $\pi_{ik}$ and $b_{ik}$: topics are always (A) present; an exchangeable (E) prior across topics on the probability that a topic is present, and a structured (S) prior on $\pi_{ik}$ where we use covariates and logistic regression to model topic presence.
\\ \indent
Topics can be always (A) present at the web site level such that $\pi_{ik} \equiv 1$ and therefore $b_{ik} \equiv 1$ for all web sites $i$ and topics $k$. In the exchangeable (E) prior, all websites have probability $\pi_{ik} \equiv \pi_k$ with $\pi_k|d_{\pi}, e_{\pi} \sim \betadist(d_{\pi}, e_{\pi})$ prior on $\pi_k$ and the $\pi_k$'s are exchangeable. We parameterize the Beta prior parameters $d_{\pi} \equiv d_{\pi}(\mu_{\pi}, \sigma_{\pi}^2)$ and $e_{\pi} \equiv e_{\pi}(\mu_{\pi}, \sigma_{\pi}^2)$ in terms of the mean $\mu_{\pi} = d_{\pi}/(d_{\pi} + e_{\pi})$ and variance $\sigma_{\pi}^2 = \mu_{\pi}(1-\mu_{\pi})/(d_{\pi} + e_{\pi} + 1)$ of $\betadist(d_{\pi}, e_{\pi})$ and place beta priors on the new parameters
\begin{align}
\mu_{\pi} &\sim \betadist(d_{\mu \pi}, e_{\mu \pi}) \label{priormeanpi} \\ 
\sigma_{\pi}^2 &\sim \betadist(d_{\sigma \pi}, e_{\sigma \pi}) \label{priorsdpi}
\end{align}
where $d_{\mu \pi}$, $e_{\mu \pi}$, $d_{\sigma \pi}$, and $e_{\sigma \pi}$ are fixed hyperparameters. 
\\ \indent
The structured (S) prior models $\pi_{ik}$ as functions of $Q$ web site covariates $X_i = (X_{i1}, \ldots, X_{iQ})'$ for web site $i$ including the intercept and let $\beta_k = (\beta_{k1}, \ldots, \beta_{kQ})'$ be the $Q$-vector of regression coefficients for topic $k$. The structured prior sets $\pi_{ik} = g(X_i' \beta_k) = \exp(X_i' \beta_k)/(1 + \exp(X_i' \beta_k))$ where $g(a) =\exp(a)/(1+\exp(a))$ is the inverse logit link.
We place a $\normaldist(\beta_0,\Sigma)$ prior on $\beta_k$, where $\beta_0$ is a prior mean $Q$-vector and $\Sigma_{Q \times Q}$ is the prior covariance matrix. We place a $\normaldist(\mu_{0},\sigma_0^2I_Q)$ prior on $\beta_0$, where $I_Q$ is the $Q$-dimension identity matrix, $\sigma_0$ is a scalar, and $\mu_0 = (\mu_{01}, \ldots, \mu_{0Q})'$ is a mean vector of length $Q$. We let $\Sigma$ be a diagonal covariance matrix with diagonal elements $\sigma_1^2, \ldots, \sigma_Q^2$ indexed by $q = 1, \ldots, Q$ and place a $\gammadist(d_{\sigma},e_{\sigma})$ prior on $1/\sigma_q^{2}$. Hyperparameters $\mu_{0}$, $\sigma_0^2$, $d_{\sigma}$, and $e_{\sigma}$ are fixed.
\\ \indent
We can similarly apply the same (A), (E), and (S) prior specifications at the web page level. Web page topic presence $c_{ijk}$ is given a $\Bernoulli(\eta_{ijk})$ prior, where $\eta_{ijk}$ is the probability of topic $k$ being present in web page $j$ of web site $i$. Topic
indicator could be always present at the web page level such that $\eta_{ijk} \equiv 1$ and $c_{ijk} \equiv 1$ for all web pages $j$, web sites $i$, and topics $k$. The exchangeable prior sets $\eta_{ijk} = \eta_k$ for all web pages $j$ and web sites $i$ and prior $\eta_k | \mu_{\eta}, \sigma^2_{\eta} \sim \betadist(d_{\eta}, e_{\eta} )$ and as at the web site level, we reparameterize in terms of the mean $\mu_{\eta} = d_{\eta}/(d_{\eta} + e_{\eta})$ and variance $\sigma_{\eta}^2 = \mu_{\eta} (1-\mu_{\eta}) / (d_{\eta} + e_{\eta} + 1)$ and set priors $\mu_{\eta} \equiv \mu_{\eta}(\mu_{\eta}, \sigma_{\eta}^2)  \sim \Betadist(d_{\mu \eta}, e_{\mu \eta})$ and $\sigma_{\eta}^2 \equiv \sigma^2_{\eta}(\mu_{\eta}, \sigma_{\eta}^2) \sim \Betadist(d_{\sigma \eta}, e_{\sigma \eta})$ and $d_{\mu \eta}$, $e_{\mu \eta}$, $d_{\sigma \eta}$, and $e_{\sigma \eta}$ are fixed hyperparameters. If we are interested in web page covariate effects, we can place a structured prior on web page topic presence. However, web page covariates are likely to be less available than web site covariates, or web page covariates may be the same as web site covariates. The health departments web site data only has web site covariates. Thus we place structured priors at the web site level only and do not consider structured priors for web page topic presence further.
\\ \indent
We thus consider six combinations of web site and web page topic presence models denoted by a two letter sequence: AA, EA, SA, AE, EE, SE, with first letter denoting the web site topic presence model, A, E, or S and the second letter denoting the web page topic presence model, A or E. In our model naming, we add local topics to these models and indicate the addition with the addition -LT.

\subsection{Gibbs Sampling}
We describe a Gibbs sampling procedure for the most complicated SE-PFA-LT model. Let a dot `$\bigcdot$' in subscripts indicate a sum across an index, for example $z_{ij\bigcdot\bigcdot}$ is the count of words in web page $j$ of web site $i$.
Let $h = 1, \ldots, z_{ij\bigcdot\bigcdot}$ index individual words in web page $j$ of web site $i$ and let $w_{ijh} \in \{1, \ldots, V\}$ and $t_{ijh} \in \{1, \ldots, K+1\}$ be the known word index and latent topic index of the $h$th word in web page $j$ of web site $i$. Let $\zeta_{ijkv} = \Phi_{ikv}\theta_{ijk} / (\sum_{k' = 1}^{K + 1}\Phi_{ik'v} 
\theta_{ijk'})$ 
be the probability of topic $k$ in web page $j$ of web site $i$ given word $v$ such that $\sum_{k=1}^{K+1}\zeta_{ijkv} = 1$. Rather than conditionally sample latent counts $z_{ijkv}$, we sample topic index $t_{ijh}$ for word $w_{ijh}$ conditional on topic weights $\theta_{ijk}$ and topic word probabilities $\Phi_{kw_{ijh}}$
\begin{align*}
t_{ijh}|\Phi_{kw_{ijh}},w_{ijh}, &\sim \text{\Multinomial}(\{\zeta_{ij1w_{ijh}},\ldots, \zeta_{ij(K+1)w_{ijh}} \})
\end{align*}
for all words in all web pages. 
Latent counts $z_{ijkv}$ at each iteration of the Gibbs sampler are deterministic functions of the $t_{ijh}$ and $w_{ijh}$. 
\\ \indent
Given the $z_{ijkv}$ and other parameters, global topic probability vector $\phi_k$, local topic probability vector $\psi_i$ and topic weight $\theta_{ijk}$ are conditionally independent and sampling is straightforward due to conjugacy with conditional densities
\begin{align*}
\phi_{k} | \{z_{\bigcdot\bigcdot kv}\}_{v} &\sim \text{\Dirichlet}(\{\alpha_{\phi} + z_{\bigcdot\bigcdot k1}, \ldots, \alpha_{\phi} + z_{\bigcdot\bigcdot kV}\}), \\
\psi_{i} | \{z_{i\bigcdot(K+1)v}\}_{v} &\sim \text{\Dirichlet}(\{\alpha_{\psi} + z_{i\bigcdot (K+1)1}, \ldots, \alpha_{\psi} + z_{i\bigcdot (K+1)V}\}), \\
\theta_{ijk}|r_k, b_{ik} = c_{ijk} = 1, z_{ijk\bigcdot} &\sim \text{\Gammadist}(r_{k}z_{ij\bigcdot\bigcdot} + z_{ijk\bigcdot}, 0.5), 
\end{align*}
and $\theta_{ijk} = 0$ if $b_{ik} = 0$ or $c_{ijk}=0$. 
\\ \indent
Conjugacy gives a convenient conditional density for the prior web page topic presence probability $\eta_k$ 
\begin{align*}
\eta_k|c_{\bigcdot\bigcdot k}, d_{\eta}, e_{\eta} &\sim \Betadist(d_{\eta} + c_{\bigcdot\bigcdot k}, e_{\eta} + N_{\bigcdot} - c_{\bigcdot\bigcdot k}).
\end{align*}
Parameters $d_{\eta}$ and $e_{\eta}$ are functions of mean $\mu_{\eta}$ and variance $\sigma_{\eta}^2$ and we use two Metropolis-Hastings \citep{mh} steps to sample $\mu_{\eta}$ and $\sigma_{\eta}^2$. 
To sample web site topic presence $b_{ik}$ and web page topic presence $c_{ijk}$, marginalize over $\theta_{ijk}$ conditional on $z_{ijk\bigcdot} = 0$ otherwise if $z_{ijk\bigcdot} > 0$ then $b_{ik} = c_{ijk} = 1$. When $z_{ijk\bigcdot} = 0$ sample
\begin{align*}
b_{ik}|\pi_{ik},r_k,\{c_{ijk}\}_j &\sim  \text{Bernoulli}\bigg(\frac{\pi_{ik}\prod_{j=1}^{N_i}(1-0.5)^{c_{ijk}r_kz_{ij\bigcdot\bigcdot}}}{1-\pi_{ik} + \pi_{ik}\prod_{j=1}^{N_i}(1-0.5)^{c_{ijk}r_kz_{ij\bigcdot\bigcdot}}}
\bigg), \\
c_{ijk}|\eta_k,b_{ik},r_k &\sim
\text{Bernoulli}\bigg(\frac{\eta_{k}(1-0.5)^{b_{ik}r_kz_{ij\bigcdot\bigcdot}}}{1-\eta_{k} + \eta_{k}(1-0.5)^{b_{ik}r_kz_{ij\bigcdot\bigcdot}}} \bigg).
\end{align*}
\\ \indent
Sampling for $r_k$, $r_0$ proceeds by introducing two families of non-negative integer-valued auxiliary random variables $\{l_{ijk}\}$ and $\{\ell_k\}$ that are conditionally distributed as the Chinese restaurant table (CRT) distribution. These auxiliary variables ensure conjugacy for sampling $r_k$ and $r_0$. The CRT has two parameters, $z$, a non-negative integer, and real valued $r > 0$. Then if $l|z,r \sim \CRT(z,r)$, $l$ has probability mass function 
\begin{align*}
P(l=\lambda|z,r) &= \frac{\Gamma(r)}{\Gamma(r + z)}|s(z,\lambda)|r^{\lambda},
\end{align*} 
where $s(\bigcdot,\bigcdot)$ denotes Stirling numbers of the first kind. Then $l$ can be sampled as a sum of independent Bernoulli random variables, $l = \sum_{m=1}^{z}y_m$, where
\begin{align*}
y_m &\sim \Bernoulli\bigg(\frac{r}{m-1+r}\bigg).
\end{align*}
Define auxiliary variables $l_{ijk}|z_{ijk\bigcdot}, r_{k}, z_{ij\bigcdot\bigcdot} \sim \CRT(z_{ijk\bigcdot}, r_{k}z_{ij\bigcdot\bigcdot})$ and $\ell_k \sim \CRT(\sum_{i=1}^{M}\sum_{j=1}^{N_i}l_{ijk}, r_{0})$ distribution. 
Then conditionally sample $r_k$ and $r_0$ as
\begin{align*}
r_{k}|\sum_{i=1}^{M}\sum_{j=1}^{N_i}l_{ijk},r_0 &\sim \text{Gamma}\bigg(r_0 + \sum_{i=1}^{M}\sum_{j=1}^{N_i}l_{ijk}, \frac{1}{1/e_r - \sum_{i=1}^{M}\sum_{j=1}^{N_i}b_{ik}c_{ijk}z_{ij\bigcdot\bigcdot}\ln(1-0.5)} \bigg) ,
\\
r_{0}|\sum_{k=1}^{K+1}\ell_{k} &\sim \text{Gamma}\bigg(d_{r0} + \sum_{k=1}^{K+1}\ell_{k}, \frac{1}{1/e_{r0} - \sum_{k=1}^{K+1}\ln(1 - u_{k})} \bigg), \\
\text{where } u_k &= \frac{-\sum_{i=1}^{M}\sum_{j=1}^{N_i}b_{ik}c_{ijk}z_{ij\bigcdot\bigcdot}\ln(1-0.5)}{1/e_r - \sum_{i=1}^{M}\sum_{j=1}^{N_i}b_{ik}c_{ijk}z_{ij\bigcdot\bigcdot}\ln(1-0.5)}.
\end{align*}
%
%
%
Sampling for $\beta_k$ conditions on auxiliary P\'{o}lya-Gamma (PG) random variables
$\{\omega_{ik}\}$. This augmentation step ensures conjugacy for sampling $\beta_k$.
Let $\omega \sim PG(b,c)$, then we can express $\omega$ as an infinite sum of independent $\Gammadist(b,1)$ variables $g_m$, such that
\begin{align*}
\omega &\overset{D}{=} \frac{1}{2\pi^2}\sum_{m=1}^{\infty}\frac{g_m}{(m-1/2)^2 + c^2/(4\pi^2)}.
\end{align*}
We approximate samples from the P\'{o}lya-Gamma distribution as a truncated sum of Gamma variables. 
\citet{weiftm} uses a P\'{o}lya-Gamma augmentation step with a truncation level of 20 to sample coefficients in modeling word presence in topics. We find that this truncation level also works well for our topic presence models with structured priors.
We introduce auxiliary variable $\omega_{ik} | X_i'\beta_k \sim \PG(1, X_i'\beta_k)$ and conditionally sample $\beta_k$ by
\begin{align*}
\beta_k|\{\omega_{ik} \}_i,\beta_{0},\Sigma &\sim \text{Normal}(\mu_k^*,\Sigma_k^*), \\
\text{where } \Sigma_k^* &= (X'\text{diag}(\{\omega_{ik} \}_i)X + \Sigma^{-1})^{-1},  \\
\mu_k^* &= \Sigma_k^*(X'\kappa_k^* + \Sigma^{-1}\beta_0),\\
\kappa_k^* &= \{b_{1k}-0.5,\ldots,b_{Mk}-0.5 \}.
\end{align*}
Prior mean coefficient vector $\beta_{0}$ has a conditional Normal posterior distribution and prior precision $\sigma_q^{2}$ has a conditional Gamma posterior distribution
\begin{align*}
\beta_{0q}|\{\beta_{kq}\}_k, &\sim \text{Normal}\bigg(\frac{(1/\sigma_0)\mu_{0q} + (1/\sigma_q)\sum_{k=1}^{K}\beta_{kq}}{1/\sigma_0 + K/\sigma_q}, (1/\sigma_0 + K/\sigma_q)^{-2}  \bigg),
\\
\sigma_q^2|\{\beta_{kp}\}_k,\mu_{0q} &\sim \text{Gamma}\bigg(d_{\sigma} + K/2, (1/e_{\sigma}+\sum_{k=1}^{K}(\mu_{0q} - \beta_{kq})^2/2)^{-1}\bigg).
\end{align*}

\subsection{Model Evaluation}
We randomly select 80\% of words in each web page to be our training set and hold out the remaining 20\% to evaluate our models. 
We keep 1000 samples after a burn in of 10,000 samples to calculate perplexity. Let superscript $s = 1, \ldots, S$ index Gibbs samples from the posterior and let $y_{i'j'\cdot v}$ be the count of held-out words $v$ in web page $j'$ in web site $i'$. We define perplexity, the log predictive probability, as 
\begin{align*}
\text{Perplexity} &= \exp(-\frac{1}{y_{i'j'\cdot v}}\sum_{i'=1}^{M'}\sum_{j'=1}^{N'_{i'}}y_{i'j'\cdot v}\log f_{i'j'v})
\intertext{where} f_{i'j'v} &= \frac{\sum_{s=1}^{S}\sum_{k=1}^{K}{\phi}_{kv}^{(s)}\theta_{i'j'k}^{(s)} + \psi_{i'}^{(s)} \theta_{i'j'(K+1)}^{(s)}}{\sum_{s=1}^{S}\sum_{k=1}^{K}\sum_{v=1}^{V}{\phi}_{kv}^{(s)} \theta_{i'j'k}^{(s)} + \psi_{i'}^{(s)}\theta_{i'j'(K+1)}^{(s)}}
\end{align*}
is the predicted probability of word $v$ of web page $j$ in web site $i$.
We repeat this random partitioning, MCMC sampling, and perplexity calculation for 5 cross validation sets and average over the 5 perplexity values for a given model. 

\section{Analyzing Web Content of Local Health Department Web Sites}
\label{analyzingwebcontent}
We analyze text data from local health department (LHD) web sites in the United States listed on the National Association of City and County Health Officials directory. Only web sites whose web address contain the text string `health', `hd', or `ph' were included. We restrict our analysis to small web sites defined as having at most 100 web pages where each web page has from 50 to at most 1000 words. 
We do not scrape web pages that are files such as .doc or .pdf files, which are often forms to be filled out. We are more interested in what is intended for people to read while browsing the web.
There are 108 LHD web sites that meet this criteria. We scraped websites for textual content using Python and Scrapy in April 2020.  We remove text items that occur on nearly every page of a web site, such as titles or navigation menus.
Common English stop words, such as `the', `and', `them', and non-alphabet characters are removed, and words are \textit{stemmed}, e.g.\ `coughing' and `coughs' are reduced to `cough'. Uncommon words defined as words occurring in fewer than 20 web pages, are removed.
The dataset analyzed has 1,061,926 total words with $V=$ 3,544 unique words across 5,863 web pages.
\\ \indent
We include a web site region covariate that indicates whether a LHD is from a state in the  Northeast, South, Midwest, or West. There are fewer than 10 LHD in either Northeast (8) and West (5) regions, and therefore we combined them into a new Northeast/West region. There are $Q=3$ web site level covariates. There are 70 web sites from LHD in Midwest states, 25 web sites from LHDs in Southern states, and 13 web sites from LHDs in either Western or Northeastern states.
We set $X_i = \{1,0,0\}$ to indicate that web site $i$ is from the Midwest region. Similarly, $X_i = \{0,1,0\}$ and $X_i = \{0,0,1\}$ indicates web site $i$ is from the South or Northeast/West region.
Coefficients $\beta_{k1}, \beta_{k2},$ and $\beta_{k3}$ correspond to intercepts for the Midwest, South, and West regions respectively. Given this specification for covariates, we are interested in the differences between regions or $\beta_{k1} - \beta_{k2}$, $\beta_{k1} - \beta_{k3}$, and $\beta_{k2} - \beta_{k3}$ for global topics $k$.

\subsection{Prior Specifications}
We model web pages nested in local health department web sites with 5 topic presence models, EA-PFA-LT, AE-PFA-LT, EE-PFA-LT, SA-PFA-LT, and SE-PFA-LT and compare it to a reference AA-PFA-LT model where topics are always present at both web page and web site levels.
We use the same hyperparameters in all models.
We choose prior for shape parameter $r_0$ such that $d_{r0} = 0.01$ and $e_{r0} = 1/.01$.
We choose priors for topic word probability vectors $\phi_{k}$ and $\psi_{k}$ such that $\alpha_{\phi} = 0.05$ and $\alpha_{\phi} = 0.05$ to encourage topics to place small probability on most words and large probability on a few words.
We set coefficient hyperparameters $\mu_0 = (0,0,0)'$, $\sigma_0 = 0.5$, $d_{\sigma} = 1$, and $e_{\sigma} = 1$ in centering the prior at the prior belief that there are no region effects and picking a prior variance that supports that a typical global topic is neither present in nearly all web sites nor unique to one web site but rather somewhere in between.
This is reflected in our specifications for the exchangeable prior on web site topic presence in EA-TPM-LT.
We specify a prior Beta($d_{\mu\pi} = 10$, $e_{\mu\pi} = 10$) prior on the prior mean of global web site topic probability $\pi_k$ and we specify a prior  Beta($d_{\sigma\pi} = 1$, $e_{\sigma\pi} = 5$) prior on the prior variance of global web site topic probability $\pi_k$.
We set hyperparameters for page topic presence probability $\eta_k$ such that $d_{\mu\eta} = 1$, $e_{\mu\eta} = K-1$, $d_{\sigma\eta} = 1$, $e_{\sigma\eta} = K-1$ in our analysis as we expect most web pages to have one or a few topics present.

\subsection{Model Comparisons}
\begin{figure}
	\includegraphics[scale = .55]{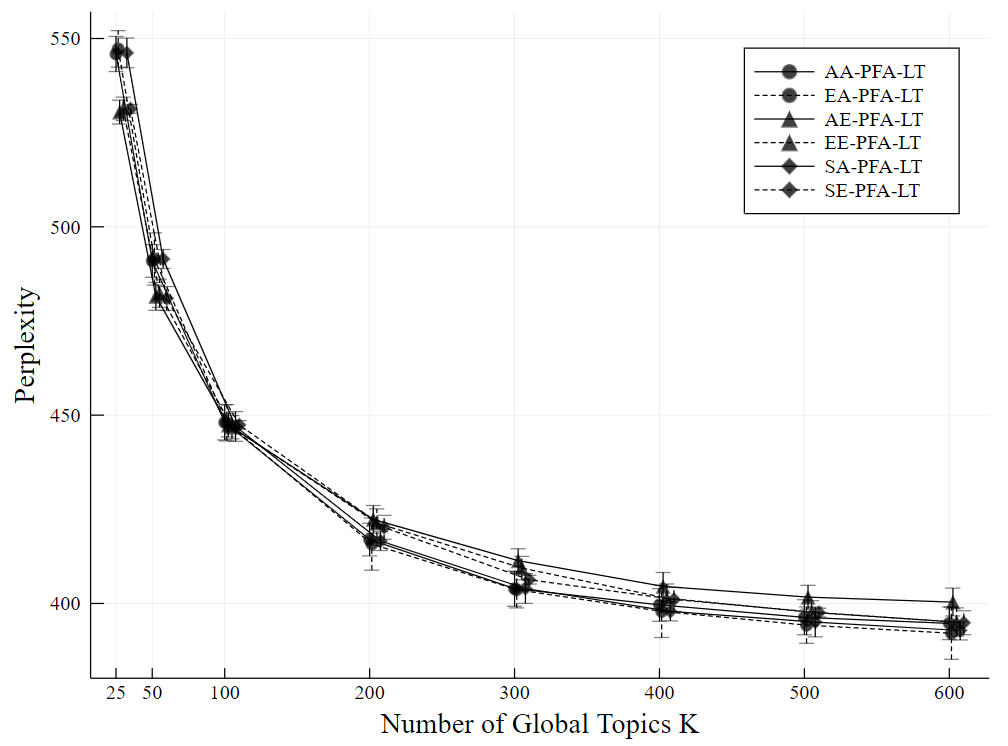}
	\caption{Perplexity by number of global topics $K$ comparison between models (lower is better).}
	\label{fig:localmodels}
\end{figure}

\begin{figure}
	\includegraphics[scale = .55]{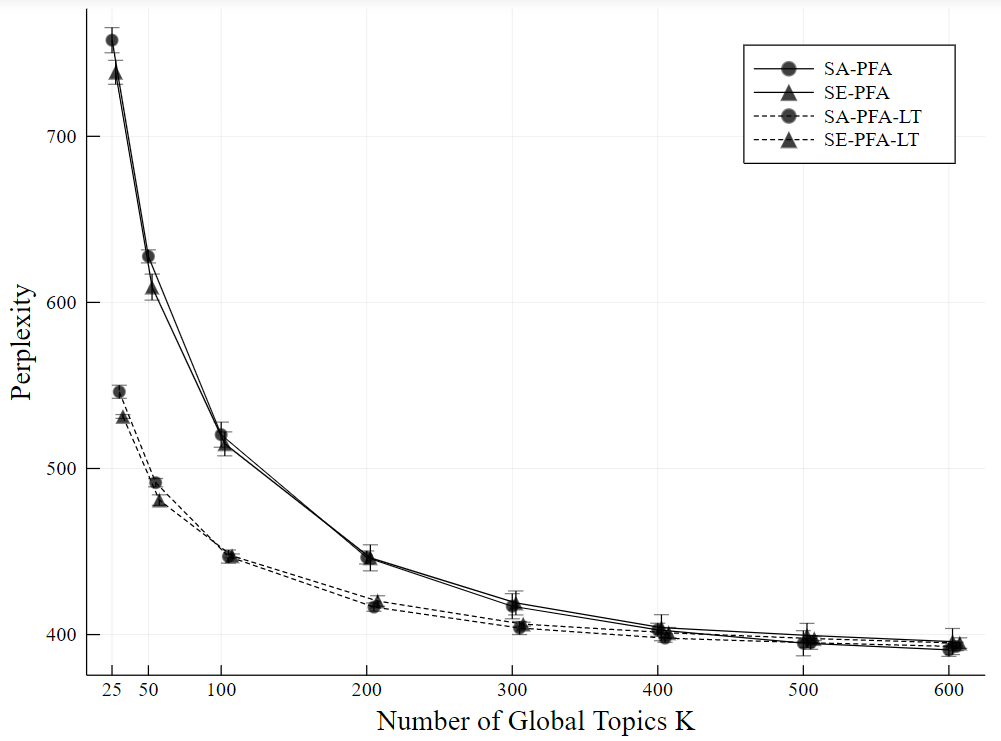}
	\caption{Perplexity comparison between topic presence models with a structured prior on web site topic presence.}
	\label{fig:localvsnolocal}
\end{figure}

We compare our models at $K=25,50,100,200,300,400,500,600$.
We further compare SA-PFA-LT and SE-PFA-LT with their no local topic counterparts SA-PFA and SE-PFA.
Figure \ref{fig:localmodels} plots the average held-out perplexity at different number of global topics $K$ for all six models with local topics.
All models perform similarly with AE-PFA-LT performing slightly worse overall.
Perplexity of all models continue to improve at $K=600$ however the difference between perplexity at $K=500$ and $K=600$ is less than 2. Further increasing $K$ increases computation time and may only improve the fit slightly. 
We model our full data with $K=500$ global topics in our analysis.
Figure \ref{fig:localvsnolocal} compares perplexity between SA-PFA-LT and SE-PFA-LT against their no local topic counterparts, SA-PFA and SE-PFA.
Models without local topics require more global topics to perform as well as models with local topics. The four models begin to perform similarly at $K=400$, where all models begin to show little perplexity improvement for each 100 increase in $K$.
We model our data with local topics in the analysis as we do not want to model covariate effects of local topics.

\subsection{Analysis of Regional Effects}
We consider the regional effects modeled with SA-PFA-LT as we are mainly interested in web site topic presence and do not want to model covariate effects of local topics. Table \ref{tab:localtopics} shows the 5 most probable words in 10 local topics. Nearly all local topics include geographical names among the 5 most probable words. 
Other high probability words in local topics are those that occur in news bulletins or other text that appears in multiple web pages of a web site.
We choose a subset of global topics from the $K=500$ global topics to review. The topics in the subset must meet three criteria, significance, frequency, and being a health topic.
First, we are interested in whether topic presence differs between regions or whether
$\beta_{kq} - \beta_{kq'}$
is significantly positive or negative for global topics $k$ and separate regions $q$ and $q'$.
Differences are significant when the 95\% sampling interval is all positive or all negative.
Second, The topic must be present in at least 20 web sites and present in at most 88 of $M = 108$ web site.
Third, the topic must be health related.
There are 101 topics that meet the significance criteria, 75 topics that further meet the frequency criteria, and 45 topics that meet all three criteria.
We select 5 topics to review.
We label them in Table \ref{tab:globaltopics} and show their 10 most probable words and the posterior mean (95\% posterior interval) of their total web site presence $b_{\cdot k}$.
\\ \indent
We carefully label each topic to avoid confusion when two topics are similar.
Similar or related topics may share common most probable words.
There were no topics that shared a similar set of most probable words with the tickborne diseases topic or the foodborn illness topic. There were two topics that are related to the CDC guidance topic; a general CDC topic with most probable words \textit{prevent, diseas, control, center, cdc, protect, reduc, accord, main, measur} and a CDC web links topic with most probable words \textit{http, wwwcdcgov, indexhtml, pdf, link, htm, indexhtm, indexphp, ncov, imag}.
The WIC nutrition and breastfeeding topics are similar in that both are related to childcare. However, the WIC nutrition topic is specifically about the WIC nutrition program while the breastfeeding topic is specifcally about breastfeeding. There is a third related mother/pregnant women topic with most probable words \textit{women, pregnanc, pregnant, infant, prenat, birth, matern, babi, mother, outcom}. The pregnancy topic does not have most probable words for nutrition or breastfeeding.
\\ \indent
Table \ref{tab:coveffects} summarizes the covariate effects in these 5 topics. The estimates are averaged over MCMC samples and intervals are 95\% MCMC intervals.
The tickborne diseases topic has most probable words \textit{tick, diseas, lyme, bite, remov, deer, tickborn, skin, transmit, attach} and is present in LHD web sites in the Midwest and West/Northeast more often than they are in LHD web sites in the South.
The foodborn illness topic has most probable words \textit{ill, foodborn, noroviru, outbreak, vomit, guidelin, suspect, diarrhea, contamin, clean} and is present in LHD web sites in the Midwest and West/Northeast more often than they are in LHD web sites in the South.
The CDC guidance topic has most probable words \textit{cdc, guidanc, recommend, updat, healthcar, guidelin, faq, advisori, disinfect, worker} and is present in LHD web sites in the West/Northeast more often than they are in LHD web sites in the South. However, the difference is borderline significant with a 95\% interval of (-3.44,-0.05) comparing South to West/Northeast.
The Special Supplemental Nutrition Program for Women, Infants, and Children (WIC) nutrition topic has most probable words \textit{wic, infant, nutrit, women, breastfeed, children, elig, food, pregnant, incom} and is present in LHD web sites in the Midwest and South more often than they are in LHD web sites in the West/Northeast.
The breastfeeding topic has most probable words \textit{breastfeed, mother, support, breast, babi, peer, counselor, milk, wic, pump} and is present in LHD web sites in the Midwest more often than they are in LHD web sites in the West/Northeast.
\\ \indent
SA-PFA-LT models how covariates are associated with web site topic presence.
We want to check if our model correctly captures these web site topic presence patterns. However, doing so manually by reading through all web pages and web sites is time consuming, thus, we describe a quick automated approach to checking using the $V^{*}$ most probable words in a topic.
For topic $k$ we check the portion of web sites in a region with at least one web page with all $V^{(*)}$ most probable word.
For example, for $V^{*} = 2$, the tickborne diseases topic is present in web site $i$ if at least one page in web site $i$ contains both words \textit{tick} and \textit{lyme}.
Many topics can be described by a few most probable words.
Thus, we let $V^{*}$ be the number of words with probability greater than 0.1 in a given topic.
We confirm that the 1 or 2 most probable words among the 5 health topics we further analyze are not identical to that of any other topic.
Table \ref{tab:programcount} shows the counts and percentages of web sites containing at least one page with $V^*$ most probable words in each region.
Our model indicates that the tickborne diseases topic is more prevalent in the Midwest and West/Northeast than in the South. 
We see the same pattern in Table \ref{tab:programcount}, where 50\% (35/70) of web sites in the Midwest and 53.8\% (7/13) of web sites in the West/Northeast have at least one page with the word \textit{tick} while 44.0\% (11/25) of web sites in the South have at least one page with the word \textit{tick}.
Similarly, for the other four topics, our logistic model results reflect the quick automated check results.

\subsection{Analysis of Tickborne Diseases Topic}
The regression results from SA-PFA-LT indicate that the tickborne disease topic is more prevalent in the Midwest and West/Northeast than in the South. This is supported by our quick automated check and further supported in a 2018 CDC report of vectorborne diseases \citep{ticks}. The report showed that from 2004-2016 the states with the top quintile of reported cases of tickborne disease are from the Midwest and Northeast. We further look into the model results for the tickborne disease topic and identify web sites that are missing the topic. More formally, we search for web sites $i$ where 97.5\% or more of MCMC samples of $b_{ik}^{(s)} = 0$.
This approach finds 24 web sites missing the tickborne disease topic; 9 from the Midwest, 12 from the South, and 3 from the West/Northeast.
\\ \indent
We further check individual web sites from the 3 West/Northeast web sites.
These three web sites belong to the La Paz County Health Department in Arizona, the Cambridge Public Health Department in Massachusetts, and the Weber-Morgan Health Department in Utah.
In the web pages we collected for these three web sites, we found no web pages with the word \textit{tick}.
The La Paz County Health Department and Weber-Morgan Health Department are from the West region where tickborne disease is not as prevalent as in the Northeast.
Upon closer inspection, we found no current online web pages from La Paz County Health Department's web site related to tickborne diseases. However, we did find a web page related to mosquitos and the Zika Virus.
We found two PDF links on Weber-Morgan Health Department's web site with the word \textit{tick}.
One is a pet disaster kit checklist, and the other is a large list of reportable diseases in Utah.
These pages were not collected as they are PDF files.
There was no dedicated informational page on tickborne diseases on Cambridge Public Health Department's web site; however, we found one news article about inviting residents to participate in a tick monitoring project. At the time of web scraping, this web page was not available to scrape. We were not able to find an archive of the news article around the date of scraping in April 2020.
Given the data we collected and modeled, SA-PFA-LT correctly identified these web sites as not having the tickborne disease topic present.

\section{Discussion}
\label{discussion}
We proposed novel topic presence models with local topics to model topic presence at two different levels in a nested document collection and apply our work to a collection of web pages nested in small web sites from local health departments in the U.S. We discussed three priors that can be placed on topic presence probabilities at web sites or web pages and showed that all topic presence models perform similarly. Thus, there is no sacrifice in fit when topic presence modeling is desired.
\\ \indent
Our AE-PFA model is similar to the sparse Gamma-Gamma PFA though. However, our AE model uses an exchangeable prior on web page topic presence probabilities rather than an independent prior where the $\eta_k$s are known a priori.  Also, we include a scaling factor of $z_{ij\bigcdot\bigcdot}$ in equation (3) while other models under the PFA construction do not; including the scaling factor adjusts for different word counts in different documents.
\\ \indent
SA-PFA-LT and SE-PFA-LT model web site topic presence probabilities conditional on web site covariates. We modeled the full data set with SA-PFA-LT to make inference on health topics and inference on regional effects on web site topic presence.
Among 500 possible topics we found many health topics where there were significant regional effects and further reviewed 5 health topics.
We found that it is important to carefully label topics as some topics are related.
After checking for related topics and distinguishing between them in labeling, we made inferences on which regions were more likely than others to have one of the health topics present.
We went further and checked several web sites that were missing the tickborne disease topic. Our model correctly identified three web sites in the West/Northeast that were missing the topic. Although, one of the three web sites did have a web page related to tick monitoring news, it was not available at the time of web scraping. Our analysis is limited to what is available online at the time. The limitation is highlighted when making inference on topic presence in a specific web site, while making inference on regional patterns allows us to leverage data from multiple web sites.

\newpage
\appendix
\section*{Tables}
\begin{table*}
	\caption{Five most probable words of 10 local topics from SA-PFA-LT with $K=500$ global topics. Most local topics include a geographical name or word among its top five words.
		Multi-county*: Logan, Morgan, Phillips, Sedgwick, Washington, and Yuma counties.
	}
	\label{tab:localtopics}
	\centering
	\begin{tabular}{llll}
		\hline
		County & State & Region & Top 5 Words \\
		\hline
		Taylor & Florida & South & \textit{florida, taylor, program, environment, link}\\	
		Wakulla & Florida & South & \textit{wakulla, water, florida, control, mosquito}\\	
		Effingham & Illinois & Midwest & \textit{effingham, illinoi, test, idph, new}\\	
		Livingston & Illinois & Midwest & \textit{livingston, news, covid, current, comment}\\	
		Vermilion & Illinois & Midwest & \textit{vermilion, illinoi, cdc, resourc, click}\\	
		Shannon & Missouri & Midwest & \textit{inspect, shannon, food, center, emin}\\	
		Hocking & Ohio & Midwest & \textit{hock, ohio, program, map, safeti}\\	
		Noble & Ohio & Midwest & \textit{nobl, ohio, provid, resourc, respons}\\	
		La Paz & Arizona & West & \textit{paz, vaccin, arizona, comment, dose}\\
		Multi-county* & Colorado & West & \textit{colorado, nchd, northeast, nchdorg, morgan}\\
		\hline
	\end{tabular}	
\end{table*}

\begin{table*}
	\caption{The 10 most probable words of 5 global health topics from SA-PFA-LT with $K=500$ global topics. Abbreviations used are Center for Disease and Control Prevention (CDC) and Special Supplemental Nutrition Program for Women, Infants, and Children (WIC). The third column lists posterior mean and 95\% posterior interval of the total number of web sites with the corresponding topic present.
	}
	\label{tab:globaltopics}
	\centering
	\begin{tabular}{llr}
		\hline
		Topic & Top 10 Words & Total Presence\\
		\hline
		Tickborne diseases & \textit{tick, diseas, lyme, bite, remov,} & 47(42,54) \\&~\textit{deer, tickborn, skin, transmit, attach}&\\	
		Foodborn illness & \textit{ill, foodborn, noroviru, outbreak, vomit,} & 50(42,58) \\&~\textit{guidelin, suspect, diarrhea, contamin, clean}&\\	
		CDC guidance & \textit{cdc, guidanc, recommend, updat, healthcar,} & 73(69,77) \\&~\textit{guidelin, faq, advisori, disinfect, worker}&\\
		WIC nutrition & \textit{wic, infant, nutrit, women, breastfeed,} & 85(83,86) \\&~\textit{children, elig, food, pregnant, incom}&\\	
		Breastfeeding & \textit{breastfeed, mother, breast, support, peer,} & 73(69,79) \\&~\textit{babi, counselor, milk, pump, mom}&\\
		\hline
	\end{tabular}	
\end{table*}

\begin{table*}
	\caption{Summary of regional differences $\hat{\beta}_{kq} - \hat{\beta}_{kq'}$ on the logit scale. Estimates are averages over 1,000 MCMC samples after a burn-in of 25,000 samples.
		An * indicates covariate effect is significant (one sided) at significance level 0.025. The MW-S column indicates the regional difference between Midwest and South. The MW-W/NE column indicates the regional difference between Midwest and West/Northeast. The S-W/NE column indicates the regional difference between South and West/Northeast.
	}
	\label{tab:coveffects}
	\centering
	\begin{tabular}{lrrr}
		\hline
		Topic & MW-S & MW-W/NE & S-W/NE \\
		\hline
		Tickborne diseases & 2.29(0.93,3.83)* & -0.30(-1.57,0.91) & -2.59(-4.38,-0.72)* \\
		Foodborn illness & 2.17(0.74,3.86)* & 0.01(-1.48,1.31) & -2.16(-4.30,-0.38)* \\
		CDC guidance & 0.84(-0.17,1.85) & -0.79(-2.48,0.60) & -1.63(-3.44,-0.05)* \\
		WIC nutrition & 0.47(-0.66,1.48) & 2.18(0.94,3.47)* & 1.71(0.30,3.23)* \\
		Breastfeeding & 0.52(-0.68,1.65) & 1.43(0.15,2.68)* & 0.91(-0.53,2.56) \\
		\hline
	\end{tabular}	
\end{table*}

\begin{table*}
	\caption{Counts and percentages of web sites with at least one web page containing the $V^*$ probable words in a topic, where $V^*$ is the number of words in a topic with probability greater than 0.1.
	}
	\label{tab:programcount}
	\centering
	\begin{tabular}{lcccc}
		\hline
		Topic & Midwest & South & West/Northeast & $V^*$ \\
		&N = 70&N = 25&N = 13&\\
		\hline
		Tickborne diseases & 35(50.0\%) & 11(44.0\%) & 7(53.8\%) & 1\\
		Foodborn illness & 42(60.0\%) & 10(40.0\%) & 9(69.2\%) & 2\\
		CDC guidance & 65(92.9\%) & 21(84.0\%) & 13(100.0\%) & 1\\
		WIC nutrition & 62(88.6\%) & 19(76.0\%) & 6(46.2\%) & 1\\
		Breastfeeding & 48(68.6\%) & 12(48.0\%) & 6(46.2\%) & 2\\
		\hline
	\end{tabular}	
\end{table*}

\newpage
\section*{Figures}

\end{document}